\documentclass[twocolumn,aps,pre,superscriptaddress,floatfix,longbibliography]{revtex4-1}
\usepackage{amsmath,amsfonts,amssymb,amsthm,graphics,graphicx,epsfig,bbm}
\usepackage[colorlinks=true,citecolor=blue,linkcolor=blue,urlcolor=blue]{hyperref}

\usepackage{graphics, graphicx}
\usepackage{subfigure}
\usepackage{dcolumn}
\usepackage{bm}
\usepackage{xcolor}
\usepackage{mathtools}
\usepackage{setspace}
\usepackage{hyperref}

\usepackage{wrapfig}
\usepackage{sidecap}
\usepackage{caption}

\begin{document}

\title{Micro-swimmer locomotion and hydrodynamics in Brinkman flows}

\date{May 12, 2025}

\author{Francisca Guzm\'an-Lastra}
\email{fguzman@uchile.cl}
\affiliation{Departamento de F\'isica, Facultad de Ciencias, Universidad de Chile, Santiago ,Chile}

\author{Enkeleida Lushi}
\email{lushi@njit.edu}
\affiliation{Mathematical Sciences, New Jersey Institute of Technology, Newark, NJ, 07102, United States}

%%%%%%%%%%%%%%%%%%%%%%%%%%%
\begin{abstract}
Micro-swimmer locomotion in heterogeneous media is increasingly relevant in biological physics due to the prevalence of microorganisms in complex environments. A model for such porous media is the Brinkman fluid which accounts for a sparse matrix of stationary obstacles via a linear resistance term in the momentum equation. We investigate two models for the locomotion and the flow field generated by a swimmer in such a medium. First, we analyze a dumbbell swimmer composed of two spring-connected spheres and driven by a flagellar force and derive its exact swimming velocity as a function of the Brinkman medium resistance, showing that the swimmer monotonically slows down as the medium drag monotonically increases. In the limit of no resistance the model reduces to the classical Stokes dipole swimmer, while finite resistance introduces hydrodynamic screening that attenuates long-range interactions. Additionally, we derive an analytical expression for the far-field flow generated by a Brinkmanlet force-dipole, which can be used for propulsive point-dipole swimmer models. Remarkably, this approximation reproduces the dumbbell swimmer's flow field in the far-field regime with high accuracy. These results provide new analytical tools for understanding locomotion in complex fluids and offer foundational insights for future studies on collective behavior in active and passive suspensions within porous or structured environments.
\end{abstract}
%%%%%%%%%%%%%%%%%%%%%%%%%%%
\maketitle

Building upon pioneering works in micro-swimmer locomotion~\cite{lighthill1952squirming, blake1971spherical}, the hydrodynamics of swimming microorganisms in homogeneous Newtonian fluids has been extensively studied in the last few decades \cite{Lauga09, Bechinger16, Spagnolie23}.

A common approximation for studying the locomotion and interactions of micro-swimmers is to model them as force dipoles in the far-field limit based on the experimental observations of the fluid flow around micro-swimmers such as bacteria and algae \cite{Drescher10, Guasto10, Drescher11}. The most common models are spherical squirmers  \cite{Ishikawa06, Ishikawa08, Spagnolie12}, self-propelling dipoles \cite{gyrya2010model}, rods or elongated ellipsoids \cite{Saintillan11, Lushi13b, Lushi14, Wioland16}, spheres connected by a spring or rod \cite{HernandezOrtiz05, Hernandez-Ortiz09, dunstan2012two, Mino11, Contino15, Lushi17}, and others \cite{Bechinger16, Desai17, Spagnolie23}. 

However, in real biological and ecological environments, microorganisms often encounter complex micro-structures --- e.g. polymeric gels, mucus layers in human tracts, extracellular matrices, and marine foams --- where the standard Stokes flow assumptions fail to fully capture the surrounding medium’s mechanical response. To adapt, microorganisms have evolved diverse locomotion strategies to swim efficiently under such non-ideal, non-Newtonian, or spatially heterogeneous conditions, particularly in the absence of inertia \cite{Cai22,Spagnolie23}. Recent experiments in complex media incorporating gel beads, arrays of obstacles or sedimenting colloids have revealed a plethora of intriguing micro-swimmer dynamics \cite{Bhattacharjee19a, Dehkharghani19, Makarchuk19, Bhattacharjee21, Bhattacharjee22, Kamdar22, Kumar22}, evincing the need for improved models that go beyond Newtonian viscous flows in order to study these phenomena. New theoretical studies have considered swimmer individual locomotion \cite{Datt15,Datt17,Chen20} or collective behavior \cite{Volpe11, Bozorgi14, LiArdekani16, Stoop19, Thijssen21} in a variety of complex flows and environments, but there is still much that remains to be understood. 

A model for viscous flows through sparse stationary networks of obstacles is the \textit{Brinkman approximation}~\cite{Brinkman47, brinkman1949calculation}, a continuum model which augments the Stokes equations with an effective hydrodynamic resistance term. This term is linear in the fluid velocity, and accounts for the drag imparted by embedded obstacles and introduces a screening length scale $\nu^{-1}$, related to the permeability of the medium \cite{Brady87, Cortez10}. Studies of individual swimmer locomotion in Brinkman flows, namely squirmer and helical or undulatory swimmers \cite{Leshansky09, Sarah16, Nguyen19, Ngangulia18,Sarah20, Chen20, Liao24}, have shown that typically a swimmer slows down in Brinkman flows due to the resistance in the medium. Recent theoretical and numerical investigations have revealed that this added resistance suppresses long-range hydrodynamic interactions between micro-swimmers, thereby altering their collective dynamics, alignment, and pattern formation~\cite{Almoteri24,Almoteri25,Daddi-Moussa25}. 

Motivated by these findings, we extend the force-dipole far-field approximation to the case of Brinkman flows to study how the modified Green's functions and screened flow fields affect the swimmer propulsion and hydrodynamic interactions. Our first model is composed of two spheres connected by a spring and propelled by a flagellum, resulting in a fluid flow generated by two anti-parallel point forces applied on the spheres. We derive the {\em exact} swimmer speed in the bulk fluid and show that resistance slows it down in comparison to a Stokesian swimmer with the same propulsion. Towards a second swimmer model that is a self-propelling dipole, we derive a far-field approximation of the Brinkmanlet doublet representing a point-force dipole in Brinkman flow. We verify that all solutions recover the corresponding Stokes flow in the limit of vanishing resistance. We compare the models to each-other, and discuss the expected hydrodynamical interactions based on the decay of the fluid-flow disturbance for small and moderate resistance values.

%%%%%%%%%%%%%%%%%%%%%%%%%%%
\begin{figure*}[htpb]
\includegraphics[width=2.0\columnwidth]{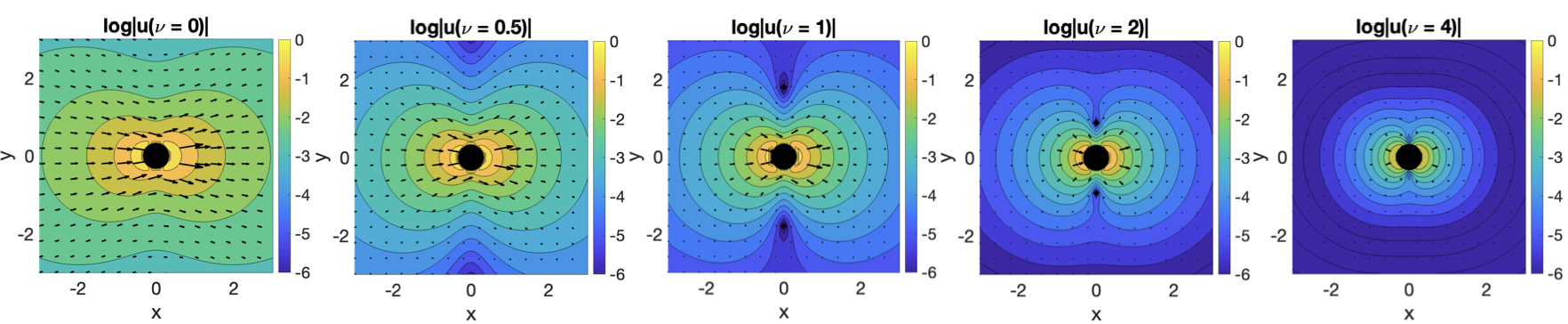}
\caption{\it The fluid flow due to force $\mathbf{g}=5(1,0,0)$ applied at the origin, for various $\nu$. The color represents $\log(|\mathbf{u}|)$. 
}
\label{fig:Flow1Brinkmanlets}
\end{figure*}

%%%%%%%%%%%%%%%%%%%%%%%%%%%
 \noindent {\bf Brinkman flow and its singularities}

The Brinkman flow in non-dimensional form is
\begin{align}\label{BrinkmanNonDim}
 \nabla^2 \mathbf{u} -\nabla q -\nu^2 \mathbf{u}= 0, \quad \quad \quad
\nabla \cdot \mathbf{u}= 0
\end{align}
where $\mathbf{u}$ is the fluid velocity, $q$ is the fluid pressure, $\nu=\ell_p/ \sqrt K_D$ is the Brinkman resistance parameter, which is also the ratio of the particle length-scale $\ell_p$, to the constant Darcy permeability length $K_D>0$ of the medium \cite{Cortez10, Vanni00}. 

The fluid flow velocity due to a point force $\mathbf{g} = g \mathbf{e}$ applied at a point $\mathbf{x}_0$ is \cite{Chwang75, Kim91, Pozrikidis92, Graham18}
\begin{align}\label{flowBrinkmanlet}
\mathbf{u}^B (\mathbf{x}) = (g/8\pi) \mathbf{B} ( \hat{\mathbf{x}}; \mathbf{e}), \quad \hat{\mathbf{x}} = \mathbf{x} - \mathbf{x}_0.
\end{align}
$\mathbf{B} ( \mathbf{x}; \mathbf{e})$ is the Green's function for the Brinkman flow, or Brinkmanlet, which in the notation of \cite{Spagnolie12}, is written as 
\begin{align}
\mathbf{B} ( \hat{\mathbf{x}}; \mathbf{e}) = A(R) \frac{\mathbf{e}}{r} + B(R) \frac{ (\hat{\mathbf{x}} \cdot \mathbf{e}) \hat{\mathbf{x}} }{r^3}
\end{align}
where $r=|\hat{\mathbf{x}}|$, $R = \nu r$, and 
\begin{align*}
A(R) &= 2 e^{-R}(1+1/R+1/R^2) -2/R^2 \\
B(R) &= -2 e^{-R}(1+3/R+3/R^2) +6/R^2.
\end{align*}
$A(0)$=$1$=$B(0)$; for $\nu$=$0$ we recover the Stokeslet $\mathbf{S} ( \hat{\mathbf{x}}; \mathbf{e})$.

Fig. \ref{fig:Flow1Brinkmanlets} shows the fluid velocity of Eq. \ref{flowBrinkmanlet} for various $\nu$.

%%%%%%%%%%%%%%%%%%%%%%%%%%%
 \noindent {\bf Dumbbell swimmer in Brinkman flow}

\begin{SCfigure}[1.1][htpb]
{\includegraphics[width=1.5in]{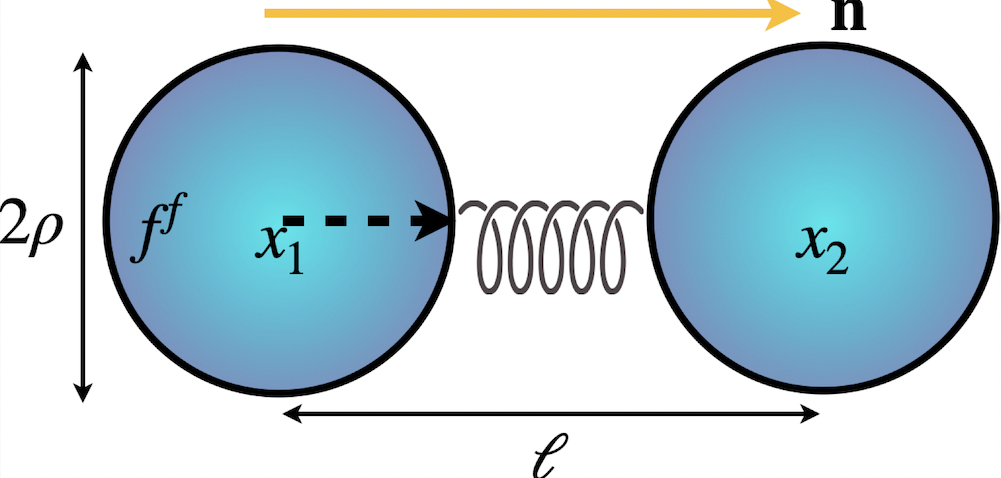}}
{\caption{\it Illustration of the dumbbell swimmer, with the ``phantom'' flagellum force indicated.  }
\label{fig:DumbbellSwimmer}}
\end{SCfigure}

Following the Stokesian swimmer model by Hernardez-Ortiz et al. \cite{HernandezOrtiz05, Hernandez-Ortiz09}, we construct our bacteria-like or pusher swimmer in Brinkman flow out of two equal-sized beads of radius $\rho$, one for the head H with center $ \mathbf{x}_2 $ and the other for the tail T with center $ \mathbf{x}_1$, connected by a stiff spring with at-rest characteristic length $\ell = |\mathbf{n}|$, where $\mathbf{n}= \mathbf{x}_2 - \mathbf{x}_1$ is the direction of propulsion. The swimmer total length and width are $\ell+2\rho$ and $2\rho$, respectively. See the illustration of the swimmer in Fig. \ref{fig:DumbbellSwimmer}.

Propulsion is provided by a ``phantom flagellum'' that, as in \cite{HernandezOrtiz05, Hernandez-Ortiz09} is not treated explicitly, but only though its effect on the swimmer body and the fluid. The tail bead which is connected to the flagellum, denoted bead 1, feels a force $\mathbf{f}^f$ exerted on it, in the direction towards the head, $\mathbf{n}= \mathbf{x}_2 - \mathbf{x}_1$, and with strength $f^f$.  The flagellum also exerts a force $-\mathbf{f}^f $ on the fluid at the position $\mathbf{x}_1$. Note that $\mathbf{f}^f$ is parallel to the swimmer direction $\mathbf{n}$.

The spring obeys a finitely non-linear elastic (FENE) model with swimmer size $\ell$ as equilibrium length:
\begin{align}
\mathbf{f}^f_c = h \frac{\mathbf{n}}{|\mathbf{n}|} \frac{ (|\mathbf{n}|-\ell) }{ 1-[ (|\mathbf{n}|-\ell)/(\ell_{max}-\ell) ]^2 }.
\end{align}
where $h=f^f/0.1 \ell$ is the spring constant, $\mathbf{n}$ the tail-to-head connector vector, and $\ell_{max}=1.15\ell$ the maximum spring length allowed, as in \cite{HernandezOrtiz05, Hernandez-Ortiz09}.

On each bead $k=1,2$, $(T,H)$ the force balance is 
\begin{equation}
\mathbf{f}^{h}_{k}+\mathbf{f}^{c}_{k}+ \delta_{k1}\mathbf{f}^{f}_{k}=0.
\end{equation}
Here $\mathbf{f}^{h}$ is the hydrodynamic drag force, $\mathbf{f}^{c}$ is the spring force, and $\mathbf{f}^{f}$ is the flagellum force. 

The hydrodynamic drag force on bead $k$ is given by the drag law for a sphere in Brinkman fluid \cite{Brinkman47, Kim91, Leshansky09}
\begin{equation}
\mathbf{f}^{h}_{k}=\zeta_B (\mathbf{v}_k - \mathbf{u}_k)
\end{equation}
for $\zeta_B=6 \pi \rho (1+\nu + \nu^2/9)$. $\mathbf{v}_k= d \mathbf{x}_k /dt$ is the velocity of the bead and $\mathbf{u}_k$ is the fluid velocity at the bead position. Note $\nu=0$ gives $\zeta=6 \pi \rho$, the Stokes drag. 

These evolution equations for the beads then are
\begin{equation}
d\mathbf{x}_k/dt= \mathbf{u}_k+ 1/\zeta_B (\mathbf{f}^{c}_{k}+ \delta_{1k}\mathbf{f}^{f}_{k}).
\end{equation}

%%%%%%%%%%%%%%%%%%%%%%%%%%%
\vspace{-0.3in}
\begin{figure}[htpb]
\includegraphics[width=0.8\columnwidth]{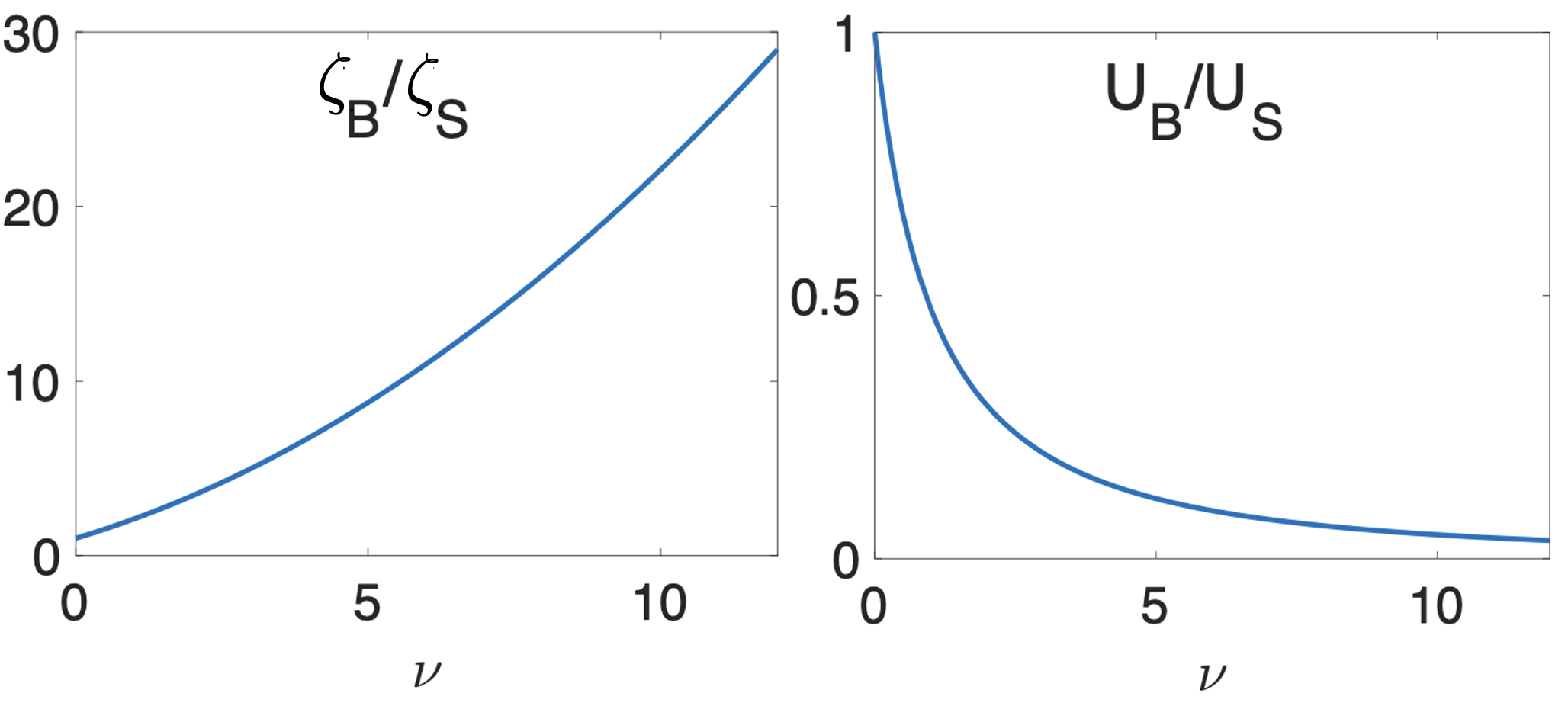}
\vspace{-0.3in}
\caption{\it The Brinkman drag $\zeta_B$ and the swimmer speed $U_B$ for various resistance values $\nu$.}
\label{fig:DragAndSpeed}
\end{figure}

%%%
\begin{figure*}[htpb]
\includegraphics[width=2.0\columnwidth]{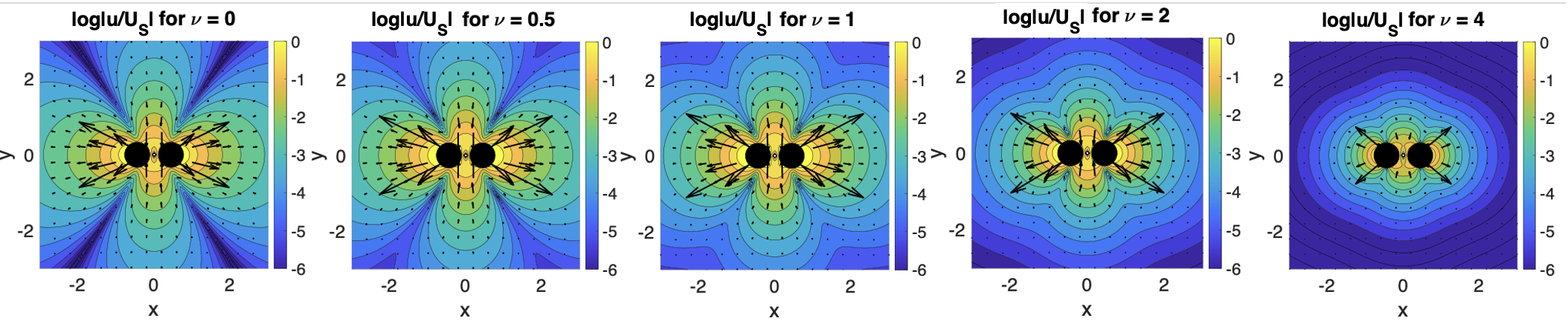}
\caption{\it The fluid flow generated by the swimmer as it moves, for various $\nu$. The color represents $\log(|\mathbf{u}|/U_S)$ %with $U_S$ being the speed of a Stokesian swimmer. 
}
\label{fig:Flow2Brinkmanlets}
\end{figure*}
%%%
\begin{figure*}[htpb]
\includegraphics[width=2.0\columnwidth]{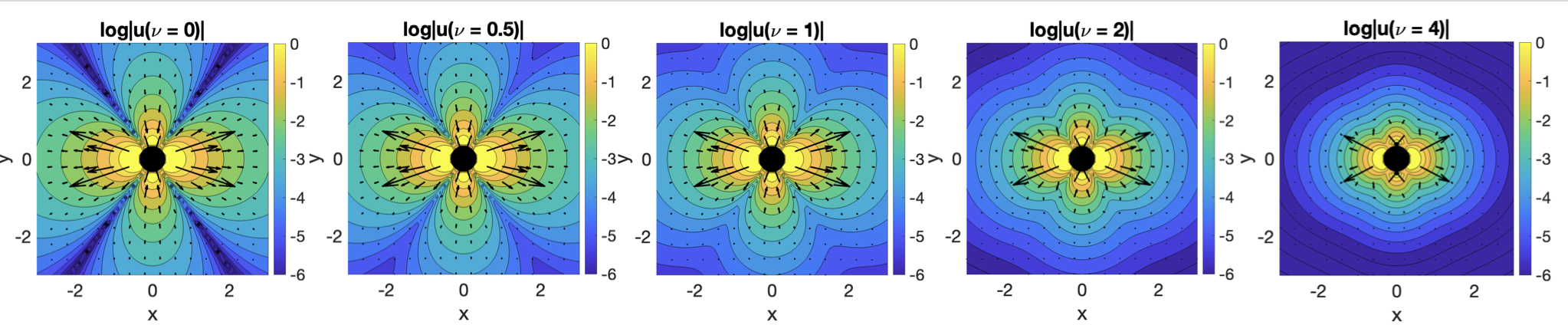}
\caption{\it The fluid flow velocity from a Brinkmanlet dipole, $\mathbf{u}^{BD}$, for various $\nu$. The color represents $\log(|\mathbf{u}|)$. 
}
\label{fig:Flow1BrinkmanDipole}
\end{figure*}

%%%
\begin{figure*}[htpb]
\includegraphics[width=2.0\columnwidth]{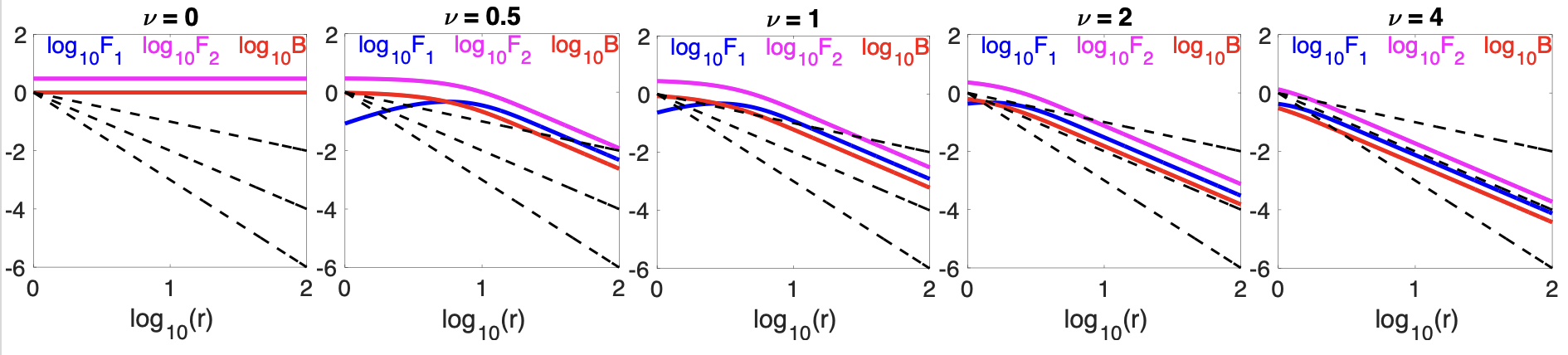}
\caption{\it Expressions $F_1$, $F_2$ and $B$ as a function of distance $r$ for various $\nu$. In dashed lines are slopes $-1, -2, -3$.
}
\label{fig:FactorsDecay}
\end{figure*}
%%%
%%%%%%%%%%%%%%%%%%%%%%%%%%%

The velocity of the fluid flow that is disturbed by this swimmer is due to the hydrodynamic force and the flagellum, $(- \mathbf{f}^{h}_{l}- \delta_{1l}\mathbf{f}^{f})$. Using the force balance equation, 
\begin{align} \label{Eq:Brinkmanswimmerflow}
\mathbf{u}(\mathbf{x})&= f_1^c \text{sign}(f_1^c)\mathbf{B}(\mathbf{x}-\mathbf{x}_1; \mathbf{e}) + f_2^c \text{sign}(f_2^c) \mathbf{B}(\mathbf{x}-\mathbf{x}_2; \mathbf{e}).
\end{align}
where $ f_1^c = |\mathbf{ f_1^c}|$ and $ f_2^c = |\mathbf{ f_2^c}|$. It follows that the the center of mass $\mathbf{x}_c= (\mathbf{x}_1 + \mathbf{x}_2)/2$ has the {\em exact} dynamics:
\begin{align}\label{xdotoneswimmer}
d\mathbf{x}_c/dt = \frac{1}{2\zeta_B} \mathbf{f}^{f}.
\end{align}

Thus the speed of this swimmer in Brinkman flow is
\begin{align}
U_B (\nu)= U_S/(1+\nu + \nu^2/9)   \label{Brinkspeeddown}
\end{align}
where $U_S$=$U_B(0)$=$f^f/(2\zeta)$ is the swimmer's speed in Stokes flow. $U_B(\nu)$ is plotted in Fig. \ref{fig:DragAndSpeed}.
%For $\nu \ll 1$, this correction in the swimmer speed is $U_B \approx U_S(1-\nu +8/9 \nu^2+...) $. 

Note that Eq. \ref{Brinkspeeddown} is an exact result, and consistent with what is found for the motion of squirmer \cite{Ngangulia18}, helical \cite{Leshansky09, Chen20} and other swimmers \cite{Sarah16, Nguyen19, Sarah20,Liao24} in Brinkman flows. It indicates that in a fluid laden with tiny obstacles and impurities a swimmer moves slower in comparison to a clean viscous fluid (Stokes) where its speed is $U_S$.

In Fig. \ref{fig:Flow2Brinkmanlets} we plot the fluid flow generated by such a swimmer, per Eq. \ref{Eq:Brinkmanswimmerflow}, for various resistances $\nu$. \\% using $f^f=$

%%%%%%%%%%%%%%%%%%%%%%%%%%%
 \noindent {\bf Brinkmanlet Dipole and fluid flow}

The far-field fluid flow for a micro-swimmer in Stokes flow can be approximated by a Stokes force dipole 
\begin{align}
\mathbf{u}^{SD} (\mathbf{x})= \alpha \mathbf{S}^D ( \hat{\mathbf{x}}; \mathbf{e}, \mathbf{e}),
\end{align}
where $\alpha$ is a constant and its sign indicates the propulsion mode (pusher or puller), and the Stokeslet dipole or doublet is 
$\mathbf{S}^D ( \hat{\mathbf{x}}; \mathbf{d}, \mathbf{e}) = \mathbf{d} \cdot \mathbf{S} ( \hat{\mathbf{x}}; \mathbf{e})$, with $\mathbf{d}$ the direction of differentiation and $\mathbf{e}$ the force-dipole and thus swimmer direction \cite{Spagnolie12}. 

We saw in Eq. \ref{Eq:Brinkmanswimmerflow} that the fluid flow disturbed by a swimmer in Brinkman flow is two anti-parallel Brinkmanlets, so it follows that the far-field fluid flow can be approximated by a Brinkmanlet force dipole
and the fluid flow resulting from it is
\begin{align}\label{eq:ubd}
\mathbf{u}^{BD}  (\mathbf{x}) = \alpha \mathbf{B}^D ( \hat{\mathbf{x}}; \mathbf{e}, \mathbf{e}).
\end{align}
The Brinkmanlet dipole or doublet is given by
\begin{align}
&\mathbf{B}^D ( \hat{\mathbf{x}}; \mathbf{d}, \mathbf{e}) = \mathbf{d} \cdot \mathbf{B} ( \hat{\mathbf{x}}; \mathbf{e}) \\
&= \left[ A(R)-A'(R)R \right] \frac{ (\mathbf{d} \cdot  \hat{\mathbf{x}}) \mathbf{e} }{r^3} - B(R) \frac{ (\mathbf{e} \cdot  \hat{\mathbf{x}}) \mathbf{d} }{r^3}   \nonumber \\
&- B(R) \frac{ (\mathbf{d} \cdot \mathbf{e})  \hat{\mathbf{x}}  }{r^3}  
+  \left[ 3B(R)-B'(R)R\right]  \frac{ (\mathbf{d} \cdot  \hat{\mathbf{x}}) (\mathbf{e} \cdot  \hat{\mathbf{x}}) \hat{\mathbf{x}} }{r^5}. \nonumber
\end{align}

In Fig. \ref{fig:Flow1BrinkmanDipole} we plot the fluid flow velocity given by Eq. \ref{eq:ubd} for strength $\alpha$ chosen to match the strength of the fluid flow in the Stokes case in Fig. \ref{fig:Flow2Brinkmanlets}. Comparing Fig. \ref{fig:Flow2Brinkmanlets} and Fig. \ref{fig:Flow1BrinkmanDipole}, we see that the two fields match well away from the swimmer, thus reinforcing the justification of using the far-field fluid flow approximation in many theoretical and computational studies of micro-swimmers in Stokes or Brinkman flows \cite{Almoteri24, Almoteri25}.

For a swimmer centered at $\mathbf{x}_0$ and $\theta$ the angle between the swimmer direction $\mathbf{e}$ and the vector $\hat{\mathbf{x}}= \mathbf{x}-\mathbf{x}_0$, the dipolar fluid flow in the Stokes case is given by \cite{Yeomans14}:
\begin{align}
\mathbf{u}^{SD} (\mathbf{x}) = \alpha \left[ 3 \cos^2 \theta -1 \right] \frac{\hat{\mathbf{x}}}{r^3}.
\end{align}
The axisymmetric dipolar fluid flow in the Stokes case decays as $r^{-2}$ with distance $r$ from the swimmer. 

In the Brinkman case, the dipolar fluid flow is:
\begin{align}
&\mathbf{u}^{BD} (\mathbf{x})/\alpha = \left[ A(R)-A'(R)R-B(R) \right] \frac{\mathbf{e}}{r^2} \cos \theta \nonumber \\
&+ \left[  (3B(R)-B'(R)R) \cos^2 \theta - B(R) \right] \frac{\hat{\mathbf{x}} }{r^3}. \label{eq:ubdtheta}
\end{align}

For low resistance values, $\nu \ll 1$,
\begin{align}%\label{eq:ubdmallnu}
\frac{\mathbf{u}^{BD} (\mathbf{x})}{\alpha}
\approx \frac{\mathbf{u}^{SD} (\mathbf{x})}{\alpha} -\frac{\nu^2}{2} \cos \theta \mathbf{e} + \frac{\nu^2}{4}{r^2} \sin^2 \theta \frac{\hat{\mathbf{x}} }{r} + \mathcal{O}(\nu r ). \nonumber
\end{align}
thus the fluid flow decays mostly as $r^{-2}$ with the resistance making minor corrections at  $\mathcal{O}(r^0 )$, and this effects can be noticed in Fig. \ref{fig:Flow1BrinkmanDipole} for $\nu=0.5$ and even $\nu=1$. 

For large resistance, $\nu \gg 1$, the far-field of the dipolar fluid flow is practically $0$ as the medium is nearly a solid. 

For intermediate resistance values $\nu$, it is difficult to explain the decay with distance of the fluid flow speed seen in Fig. \ref{fig:Flow1BrinkmanDipole} from the formula in Eq. \ref{eq:ubdtheta}, thus we look at the expressions $F_1=A(R)-A'(R)R-B(R)$ and $F_2=3B(R)-B'(R)R$ as well as $B(R)$ as a function of distance $r$ from the swimmer in logarithmic plots in Fig. \ref{fig:FactorsDecay}. The plots elucidate the behavior. For small $\nu$, $F_1~0,F_2 ~ r^0, B~r^0$ within 10 swimmer lengths, as expected. We see that $F_1~r^{-2},F_2 ~ r^{-2}, B~r^{-2}$ for distances $>$ 2 swimmer lengths for $\nu=2$, and within a swimmer length for $\nu=4$, indicating that the dipolar fluid flow decays as $r^{-4}$ for $\nu>1$.

%%%%%%%%%%%%%%%%%%%%%%%%%%%
 \noindent {\bf Summary and Discussion}

We have analytically studied the flow field generated by two micro-swimmer models in a Brinkman fluid. First, we revisited the dumbbell swimmer---a system of two spheres connected by a spring---propelling through a porous medium characterized by a resistance parameter $\nu$ and the drag coefficient $\zeta_B$. We derived the exact solution for the swimmer’s velocity in the bulk and observed that, the Brinkman drag increases monotonically with $\nu$, resulting in a reduced swimmer speed  in heterogeneous Brinkman fluid compared to the homogenous viscous flow or Stokes case. In the limit $\nu \to 0$, we recover the classical dipolar flow of the dumbbell swimmer in a Newtonian fluid, while increasing $\nu$ introduces screening effects that attenuate fluid distortions.

We  showed that the far-field flow of the dumbbell model can be approximated by a Brinkmanlet force dipole. We derived the corresponding velocity field and analyzed how the Brinkman resistance modifies the standard Stokes dipole solution. In particular, we found that the flow decays as $r^{-2}$ for small $\nu$, transitioning to a faster decay of $r^{-4}$ when $\nu > 1$.

Our findings highlight key modifications to micro-swimmer locomotion and the fluid fluid they induce in complex media. This work lays the groundwork for future studies on collective dynamics and environment-mediated interactions in Brinkman-type fluids. Lastly, out models also offer a tractable framework to study hydrodynamic interactions in active and passive suspensions in porous media and complex flows.

%%%%%%%%%%%%%%%%%%%%%%%%%%%
 \noindent {\bf Acknowledgements}
F.G.-L. acknowledges support from the ANID-Fondecyt Regular No. 1250913. E.L. acknowledges support from the Simons Foundation. The authors thank Y. Almoteri and D. Pushkin for helpful discussions, and the Newton Institute for Mathematical Sciences, Cambridge, for inclusion and hospitality during the program “Anti- diffusive dynamics: from sub-cellular to astrophysical scales” supported by EPSRC grant no. EP/R014604/1.

\bibliography{references}

\end{document}